\documentclass[pra,aps,amsmath,amssymb,showpacs,preprint]{revtex4}
\usepackage{graphicx}
\usepackage{subfigure}
\usepackage{natbib}
\usepackage{hyperref}
\usepackage{float}
\hypersetup{
    unicode=false,          
    pdftoolbar=true,        
    pdfmenubar=true,        
    pdffitwindow=false,     
    pdfstartview={FitH},    
    pdftitle={My title},    
    pdfauthor={Author},     
    pdfsubject={Subject},   
    pdfcreator={Creator},   
    pdfproducer={Producer}, 
    pdfkeywords={keywords}, 
    pdfnewwindow=true,      
    colorlinks=true,        
    linkcolor=red,          
    citecolor=blue,         
    filecolor=magenta,      
    urlcolor=cyan           
}

\begin{document}

\title{Environmental Effects on the Geometric Phase}

\author{A. C. G\"{u}nhan}
\affiliation{Department of Physics, Mersin University, 33343 Mersin, Turkey}
\author{S. Turgut}
\author{N. K. Pak}
\affiliation{Department of Physics, Middle East Technical University, 06531 Ankara, Turkey}

\date{\today}

\begin{abstract}
The behavior of the geometric phase gained by a single spin-$1/2$
nucleus immersed into a thermal or a squeezed environment is
investigated. Both the time dependence of the phase and its value
at infinity are examined against several physical parameters. It
is observed that for some intermediate ranges of the temperature
and the coupling strength, the presence of squeezing enhances the
geometric phase.
\end{abstract}


\maketitle
\section{Introduction}
\label{s1} The geometry of the Hilbert space of a quantum system
is registered to the memory of the system as a geometric phase
factor\cite{Pancharatnam56,Berry84,AharanovAnandan87,AnandanAharanov88,SamuelBhandari88,Uhlmann91,Anandan92}.
It is inherently fault tolerant because of its geometric
nature\cite{Preskill99}. For this reason, it is used in the
implementation of controlled gates for quantum
computers\cite{Ekert00,HodgesEtal07}. These gates are also of
crucial importance for a universal set of quantum logic
gates\cite{Ekert00,HodgesEtal07,NielsenChuang00}. For this reason,
a full understanding of the nature of quantum geometric phase in
different environments is vital for quantum computations.

Various aspects of the effects of the environment on the GP of
open quantum systems have been studied. Rezakhani and Zanardi
analyzed the temperature effects on mixed-state GP for a single
and two coupled spin-$1/2$ particles\cite{RezakhaniZanardi06}.
Wang \textit{et al.} analyzed the effects of a squeezed vacuum
reservoir on GP of a two-level atom in an electromagnetic field by
a formulation entirely in terms of geometric
structures\cite{WangEtal07}. Carollo \textit{et al.} showed that
GP can be induced by cyclic evolution in an adiabatically
manipulated squeezed vacuum reservoir\cite{CarolloEtal06}.
Banerjee and Srikanth studied the effects of a squeezed-thermal
environment on the GP of a two-level
system\cite{BanerjeeSrikanth08} for dissipative and
non-dissipative cases, and analyzed the initial-state and
temperature dependence of GP for the system. The purpose of this
article is to examine the effect of various physical parameters on
the time-dependence of GP.

The system analyzed in this article is a two-level nucleus inside
a magnetic field. The time evolution is non-unitary due to the
interaction with the environment. The environment is first taken
as a thermal bath. Then, this bath is considered to be driven by
an electromagnetic field in a squeezed state in order to see
whether the GP can be enhanced. The dependence of the GP on
temperature, external magnetic field, coupling strength, squeezing
and initial state of the system are then analyzed. The GP is
computed by using the kinematic definition given by Tong \emph{et
al.}\cite{TongEtal04}. In their approach, when the system's
density matrix in the Schr\"{o}dinger picture is $\hat{\rho}(t)$,
the GP gained in the time interval $[0,\tau]$ is given by the
expression
\begin{equation}\label{FAY}
 \Phi(\tau)=\arg\left(\sum_k \sqrt{\lambda_{k}(0)\lambda_{k}(\tau)} \langle\varphi_{k}(0)\vert\varphi_{k}(\tau)\rangle
     e^{(-\scriptstyle\int_{0}^{\tau} \langle\varphi_{k}(t)\vert\dot{\varphi}_{k}(t)\rangle {\tt d}t )}\right)
     ~~,
\end{equation}
where $\lambda_{k}(t)$ and $\vert\varphi_{k}(t)\rangle$ are the
eigenvalues and the eigenvectors of $\hat{\rho}(t)$, respectively.

The content of the article is as follows. In section
\ref{sec:system}, the system and its interaction with the
environment is described. This section also includes the
derivation of the non-unitary dynamics of the reduced density
matrix by using Markov approximation. After that, the density
matrix of the system is computed analytically. In section
\ref{sec:GP_dependence}, the dependence of the GP on various
physical parameters is analyzed. Finally, section \ref{sec:conc}
contains brief conclusions.

\section{The Nucleus inside a Bath}
\label{sec:system}

Our specific system is a single spin-$1/2$ nucleus in an external
static magnetic field $\vec{B}$ which is taken to be in the $z$
direction. The Hamiltonian of the nucleus is
\begin{equation}\label{1NucleusHam}
    \hat{H}_{N}=-\hbar\omega_{N}\hat{I}^{z}~~,
\end{equation}
where $\hbar$ is the Planck constant,
$\omega_{N}=\gamma_{N}|\vec{B}|$, $\gamma_{N}$ is the gyromagnetic
ratio of the nucleus, $\hat{I}^{z}=\hat{\sigma}^z/2$ and
$\hat{\sigma}^z$  is the $z-$component of the Pauli spin operator.

The environment (reservoir) is assumed to be a bath of harmonic
oscillators (such as electromagnetic radiation) where the
frequency spectrum forms a continuum. The annihilation (creation)
operator for the mode at frequency $\omega$ is denoted by
$\hat{b}(\omega)$ ($\hat{b}^\dagger(\omega)$) and the Hamiltonian
of the environment is
\begin{eqnarray}\label{ResHamil}
\hat{H}_{R}=\hbar \displaystyle\int_{0}^{\infty}  \omega
\hat{b}^{\dagger}(\omega)\hat{b}(\omega){\tt d} \omega~~.
\end{eqnarray}
The interaction of nucleus with the oscillators is described by
the interaction Hamiltonian,
\begin{equation}\label{IntHam}
    \hat{H}_{NR}=\hbar \displaystyle\int_{0}^{\infty}
    g_{N}(\omega)\hat{I}^{+}\hat{b}^{\dagger}(\omega){\tt d} \omega\, +    h.c.~~,
\end{equation}
where $g_{N}(\omega)$ is the coupling coefficient between the
nucleus and the $\omega$-mode of the reservoir,
$\hat{I}^{\pm}=\hat{I}^x\pm i\hat{I}^y$ are the spin ladder
operators, and $h.c.$ indicates the hermitian conjugate term.

The total Hamiltonian
$\hat{H}=\hat{H}_{N}+\hat{H}_{R}+\hat{H}_{NR}$, is transformed
into the interaction picture as
\begin{equation}\label{1NucleusHamIntPict1}
   \hat{V}(t)=\hbar\,\displaystyle\int_{0}^{\infty} g_{N}(\omega)e^{i(\omega-\omega_{N})t}\hat{I}^{+}\hat{b}^{\dagger}(\omega){\tt d}
   \omega\,+h.c.~~.
\end{equation}
Let $\hat{\rho}_{NR}(t)$ denote the state of the combined system
of the nucleus and the environment at time $t$, and let
$\hat{\rho}_{NR}^I(t)$, $\hat{\rho}_{R}^I(t)$ and
$\hat{\rho}_{N}^I(t)$ denote the interaction picture states of the
combined system, the environment and the nucleus, respectively. We
study this system in the regime in which the Markov approximation
can be applied. In this approximation, it is assumed that the
environment remains in its initial state during the evolution,
$\hat{\rho}_{R}^I(t)=\hat{\rho}_{R}(0)$, and the state of the
combined system is taken to be in product form,
$\hat{\rho}_{NR}^I(t)=\hat{\rho}_{N}^I(t)\otimes\hat{\rho}_{R}(0)$.
As a result, the equation of motion of the state of the nucleus
can be obtained as
\begin{equation}\label{SEqnOfMotion}
    \dot{\hat{\rho}}_{N}^I(t)=-\frac{i}{\hbar} \mathrm{tr}_{R}[\hat{V}(t),\hat{\rho}_{N}^I(0)\otimes\hat{\rho}_{R}(0)]-\frac{1}{\hbar
    ^2}\mathrm{tr}_{R}\displaystyle\int_{0}^{t}[\hat{V}(t),\;[\hat{V}(t')\;,\hat{\rho}_{N}^I(t')\otimes\hat{\rho}_{R}(0)]\;]\;{\tt d} t',
\end{equation}
where the dot denotes the time derivative and $\mathrm{tr}_{R}$
represents trace over the degrees of freedom of the
environment\cite{BarnettRadmore97,ScullyZubairy97}.

For the thermal environment, the state is given by
$\hat{\rho}_R(0)=\hat{\rho}_{\textrm{th}}=\exp(-\hat{H}_R/k_BT)/\mathrm{tr}\exp(-\hat{H}_R/k_BT)$).
If the bath is driven by a squeezed field, the state of the
squeezed-thermal environment is given by
\begin{equation}
  \hat{\rho}_R(0)= \hat{S}\hat{\rho}_{\textrm{th}}\hat{S}^\dagger,
\label{eq:squeezed_density_matrix}
\end{equation}
where
\begin{equation} \label{ContSqueezingOper}
\hat{S}=\hat{S}[\xi(\omega)]=\exp\left(
-\frac{1}{2}\displaystyle\int_{\omega}^{2\Omega} {\tt d} \omega [
\xi(\omega)\hat{b}^{\dagger}(\omega)\hat{b}^{\dagger}(2\Omega -
\omega)-\xi^{*}(\omega)\hat{b}(2\Omega - \omega)
\hat{b}(\omega)]\right),
\end{equation}
where $2\Omega$ is the squeezing-carrier frequency,
$\xi(\omega)=r(\omega)e^{i\phi(\omega)}$, $r(\omega)$ and
$\phi(\omega)$ are real numbers characterizing the squeezing which
satisfy $\xi(\omega)=\xi(2\Omega-\omega)$.

For both types of environments, we have $\langle
b(\omega)\rangle=0$ which makes the first term on the right-hand
side of Eq.~(\ref{SEqnOfMotion}) vanish. The second term can be
expressed in the form
\begin{eqnarray}
  \dot{\hat{\rho}}_{N}(t)   &=& -\frac{i}{\hbar}[\Delta \hat{H}_N,\hat{\rho}_{N}(t)]+
          C_- \left(2\hat{I}^-\hat{\rho}_{N}(t)\hat{I}^+ - \hat{\rho}_{N}(t)\hat{I}^+\hat{I}^-  -\hat{I}^+\hat{I}^-\hat{\rho}_{N}(t)\right) \nonumber \\
     &  & +C_+ \left(2\hat{I}^+\hat{\rho}_{N}(t)\hat{I}^- - \hat{\rho}_{N}(t)\hat{I}^-\hat{I}^+  -\hat{I}^-\hat{I}^+\hat{\rho}_{N}(t)\right)  \nonumber \\
     &  & - D \, \hat{I}^+\hat{\rho}_{N}(t)\hat{I}^+ - D^* \, \hat{I}^-\hat{\rho}_{N}(t)\hat{I}^-,
\label{eq:Time_Evolution}
\end{eqnarray}
where $\Delta \hat{H}_N=-\hbar\Delta\omega_N\hat{\sigma}^z/2$
represents a re-normalization of the frequency $\omega_N$ of the
nucleus, and $C_\pm$ and $D$ are numbers that capture the
collective effect of all of the oscillators in the reservoir
satisfying
\begin{equation}
  C_+ = C_- +\pi \vert g_N(\omega_N)\vert^2~~~.
\end{equation}
Each of these constants can be expressed as an integral over the
frequency $\omega$, which can be decomposed into a principal-value
integral, which depends on the precise frequency dependence of
$g_N(\omega)$, and a Dirac-delta integral term, which depends only
on $g_N(\omega_N)$. The coupling function $g_N(\omega)$ is freely
adjustable; by changing that function suitably, the values of all
principal-value integrals and therefore all final constants
$C_\pm$, $D$ and $\Delta\omega_N$ can be adjusted to any desirable
value. Here, for simplicity, all of these principal-value
integrals are taken to be zero and the resultant constants are
used. With this choice, the re-normalized Hamiltonian $\Delta
\hat{H}_N$ becomes zero.

For the purely thermal environment, the constants in
Eq.~(\ref{eq:Time_Evolution}) can be computed as $D=0$ and
$C_-=\pi\vert g_N(\omega_N)\vert^2 n(\omega_N)$, where
\begin{equation}
  n(\omega)=\frac{1}{\exp(\hbar\omega/k_BT)-1}~~
\end{equation}
denotes the average occupation number of the mode at frequency
$\omega$. In this case, the matrix elements of the interaction
picture density matrix can be integrated to
\begin{eqnarray}
  \rho_{N11}^I(t) &=& \rho_{N11}^I(\infty) +  (\rho_{N11}^I(0)-\rho_{N11}^I(\infty))e^{-2(C_++C_-)t} ~~~,
     \label{eq:Rho_Thermal_Diagonal}   \\
  \rho_{N12}^I(t) &=& \rho_{N11}^I(0) e^{-(C_++C_-)t} ~~~,
\end{eqnarray}
where
\begin{equation}
  \rho_{N11}^I(\infty)=\rho_{N11}(\infty)=\frac{C_+}{C_++C_-}~~.
\end{equation}

For the squeezed-thermal environment given in
Eq.~(\ref{eq:squeezed_density_matrix}), the constants can be found
as
\begin{eqnarray}
  C_- &=& \pi\vert g_N(\omega_N)\vert^2    \Big\{n(\omega_N)+\left[n(\omega_N)+n(2\Omega-\omega_N)+1\right]\sinh^2 r(\omega_N)\Big\}~,\\
  D &=& \pi g_N(\omega_N)g_N(2\Omega-\omega_N) \left[n(\omega_N)+n(2\Omega-\omega_N)+1\right]\sinh (2r(\omega_N))
            e^{2i(\Omega-\omega_N)t-i\phi(\omega_N)}.
\end{eqnarray}
For this case too, the time-dependent density matrix can be found
analytically. The diagonal entry $\rho_{N11}^I(t)$ is still given
by Eq.~(\ref{eq:Rho_Thermal_Diagonal}) and the off-diagonal entry
can be expressed in general as
\begin{equation}
  \rho_{N12}^I(t) = \left(A_1 e^{st}+A_2 e^{-st}\right)e^{-(C_++C_-)t+i(\Omega-\omega_N)t}~~,
  \label{eq:rho_12_t_dep}
\end{equation}
where $s$ is the purely real or purely imaginary constant
\begin{equation}
 s=\sqrt{\vert D\vert^2-(\Omega-\omega_N)^2},
\end{equation}
and $A_1$ and $A_2$ are constants that should be determined from
the initial state $\hat{\rho}_N(0)$. The squeezing changes the
time-dependence of the density matrix as follows. First, it
changes the long-time limit of the density matrix
$\hat{\rho}_N(\infty)$. It also makes the diagonal relaxation
time, $(2(C_++C_-))^{-1}$, shorter. Apart from those, it changes
the time-dependence of the off-diagonal entry; in particular it
produces new oscillatory behavior for sufficiently large
squeezing.

The geometric phase $\Phi(\tau)$ given in Eq.~(\ref{FAY}) can be
computed analytically in the thermal case. The argument of the
exponential in this expression is given by
\begin{equation}
 -\displaystyle\int_{0}^{\tau} \langle\varphi_{\pm}(t)\vert\dot{\varphi}_{\pm}(t)\rangle {\tt d}t
    =\mp i \frac{\omega_N}{2}\left(\tau +
        \frac{1}{C_++C_-}(F(\tau)-F(0))\right)
\end{equation}
where
\begin{eqnarray}
  F(\tau) &=& \ln(A_\tau -\delta_\tau+2\delta_\infty)
              -\ln\left(   2(\delta_0-\delta_\infty) + \frac{\vert \rho_{N12}(0)\vert^2}{A_\tau+\delta_\tau}  \right)~~,  \\
  A_\tau &=& \sqrt{\frac{1}{4}-\det \hat{\rho}_N(\tau)} ~~,\\
  \delta_\tau &=&  \frac{1}{2}\left(\rho_{N11}(\tau)-\rho_{N22}(\tau)\right)~~,
\end{eqnarray}
and $\vert\varphi_\pm(\tau)\rangle$ are eigenvectors with
corresponding eigenvalues $\lambda_\pm(\tau)=1/2\pm A_\tau$.

 It should be noted that when $\tau$ goes to infinity, the
density-matrix $\hat{\rho}_N(\tau)$ goes to a diagonal
time-independent state. Hence $F(\tau)$ and consequently the phase
$\Phi(\tau)$ tend to constant limits. For the squeezed-thermal
environment, the argument of the exponential needs to be computed
numerically but it can be shown that the same behavior holds at
infinity, i.e., the GP settles down to a well-defined finite
limit.

\section{The Dependence of the Geometric Phase on Physical Parameters}
\label{sec:GP_dependence}

For investigating the behavior of the GP under different physical
conditions, the principal-value integrals are taken to be zero as
explained in the previous section, and both of the coupling
coefficients $g_N(\omega_N)$ and $g_N(2\Omega-\omega_N)$ are taken
to be equal to a real positive value $g$. The phase parameter
$\phi(\omega_N)$ of the squeezed state is taken to be $0$. The
relevant parameter $r(\omega_N)$ that gives the amount of
squeezing is simply denoted by $r$ below.

In the problem, there are three parameters having the dimension of
energy: the excitation energy $\hbar\omega_N$ (which is
proportional to the external magnetic field), the thermal energy
$k_BT$ and an energy related to the coupling strength $\hbar g^2$.
The behavior of the GP is invariant if all three of these
parameters are scaled by the same amount. For this reason, only
the ratios of these parameters need to be specified as done below.

\subsection{Time Dependence of the Geometric Phase}

The dependence of the geometric phase $\Phi(\tau)$ on time $\tau$
has periodic structures with period equal to $2\pi/\omega_N$. In
each of the oscillations over one period, $\Phi(\tau)$ also
changes by an amount depending crucially on the values of all
physical parameters. If there is squeezing, an additional periodic
structure coming from the time-dependence of the off-diagonal
entry given in Eq.~(\ref{eq:rho_12_t_dep}) may appear.

\begin{figure}[h!]
\includegraphics[scale=0.42]{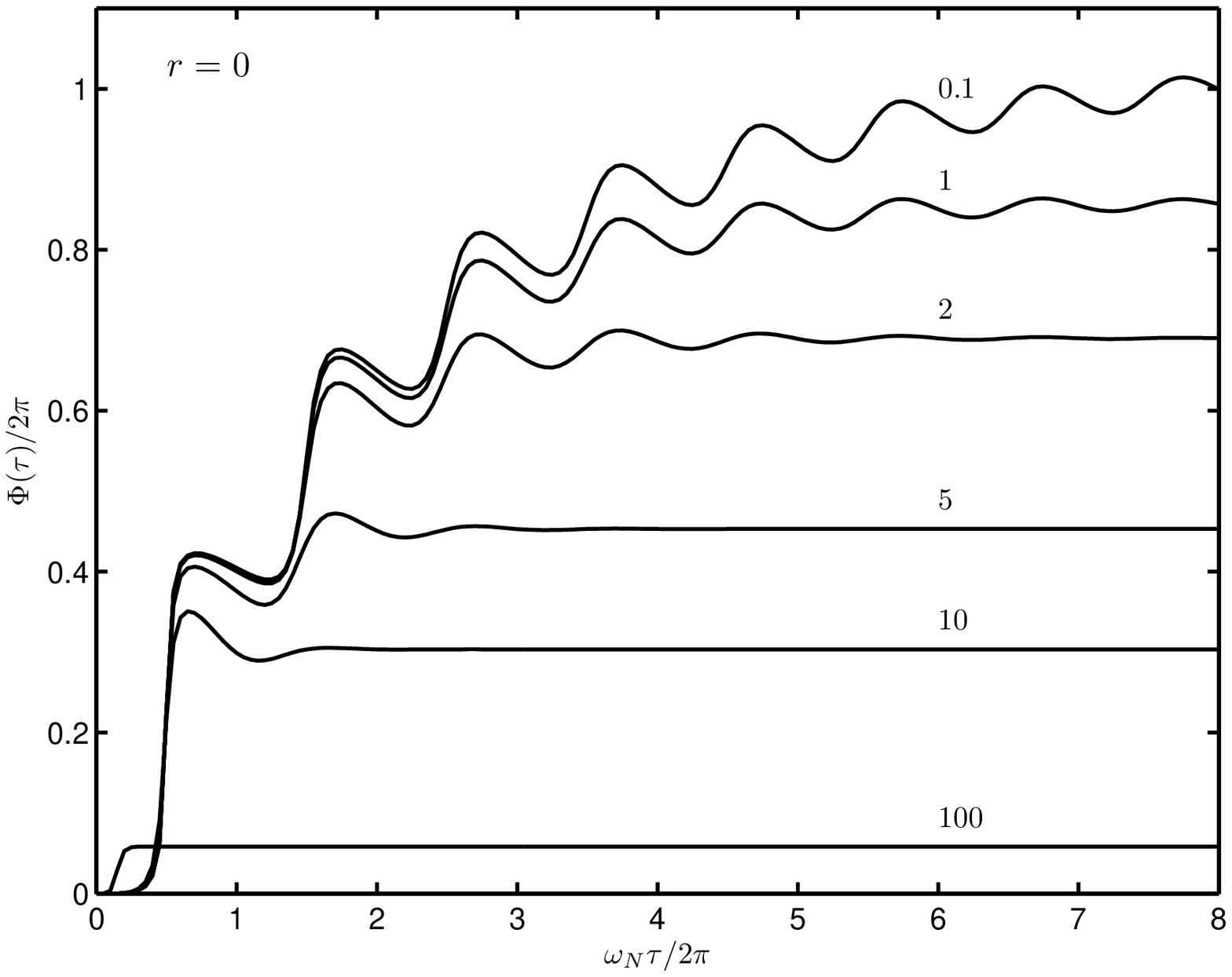}
\includegraphics[scale=0.42]{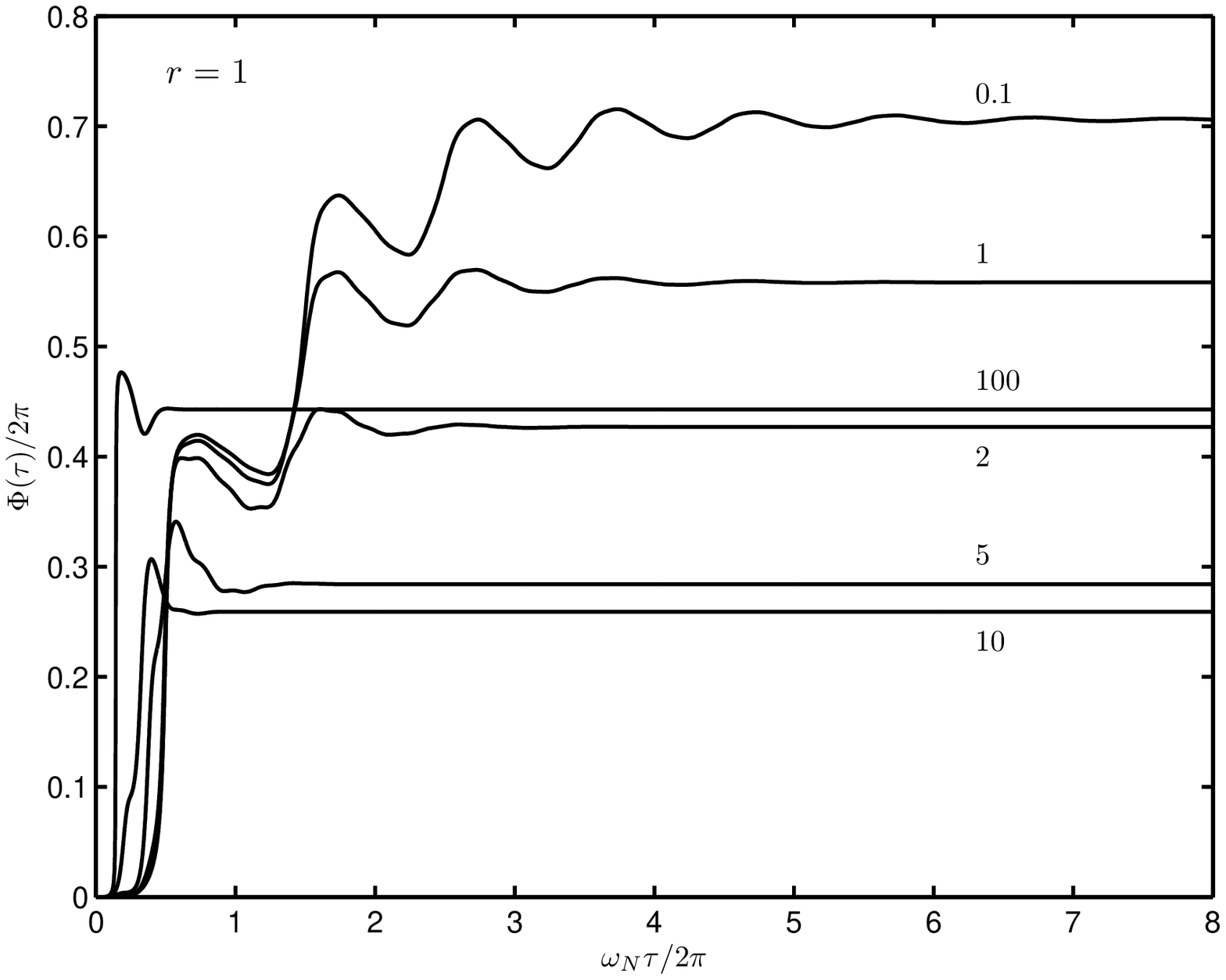}
\caption{The GP vs time for different temperatures for a thermal
environment (left) and a squeezed-thermal environment with $r=1$
(right). The coupling strength is $g^2/\omega_N =0.01$ and the
initial state is a pure spin up state along $x$. The numbers on the
graph indicate the value of $k_BT/\hbar\omega_N$.}
\label{fig:sq_temp}
\end{figure}

\begin{figure}[h!]
\includegraphics[scale=0.42]{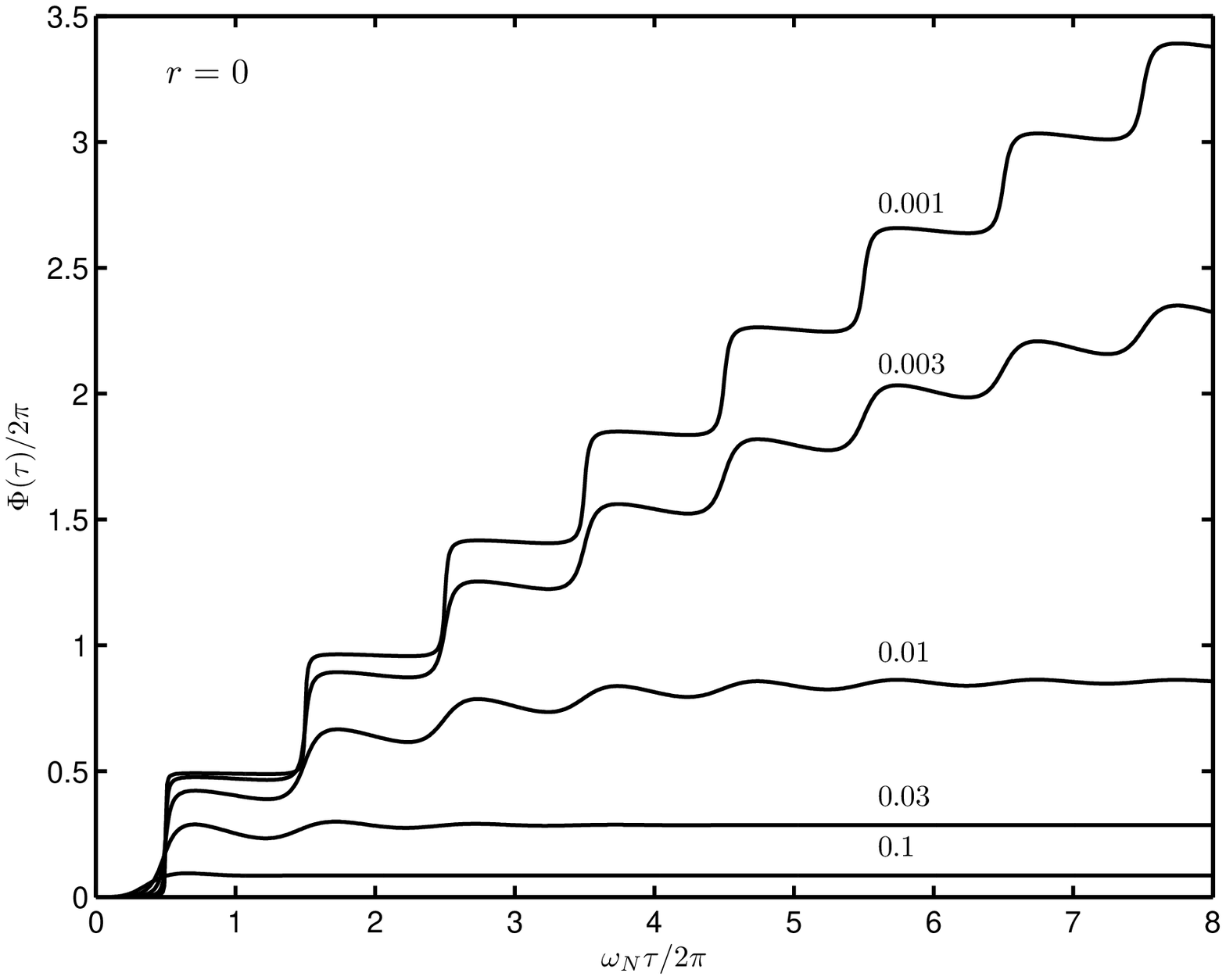}
\includegraphics[scale=0.42]{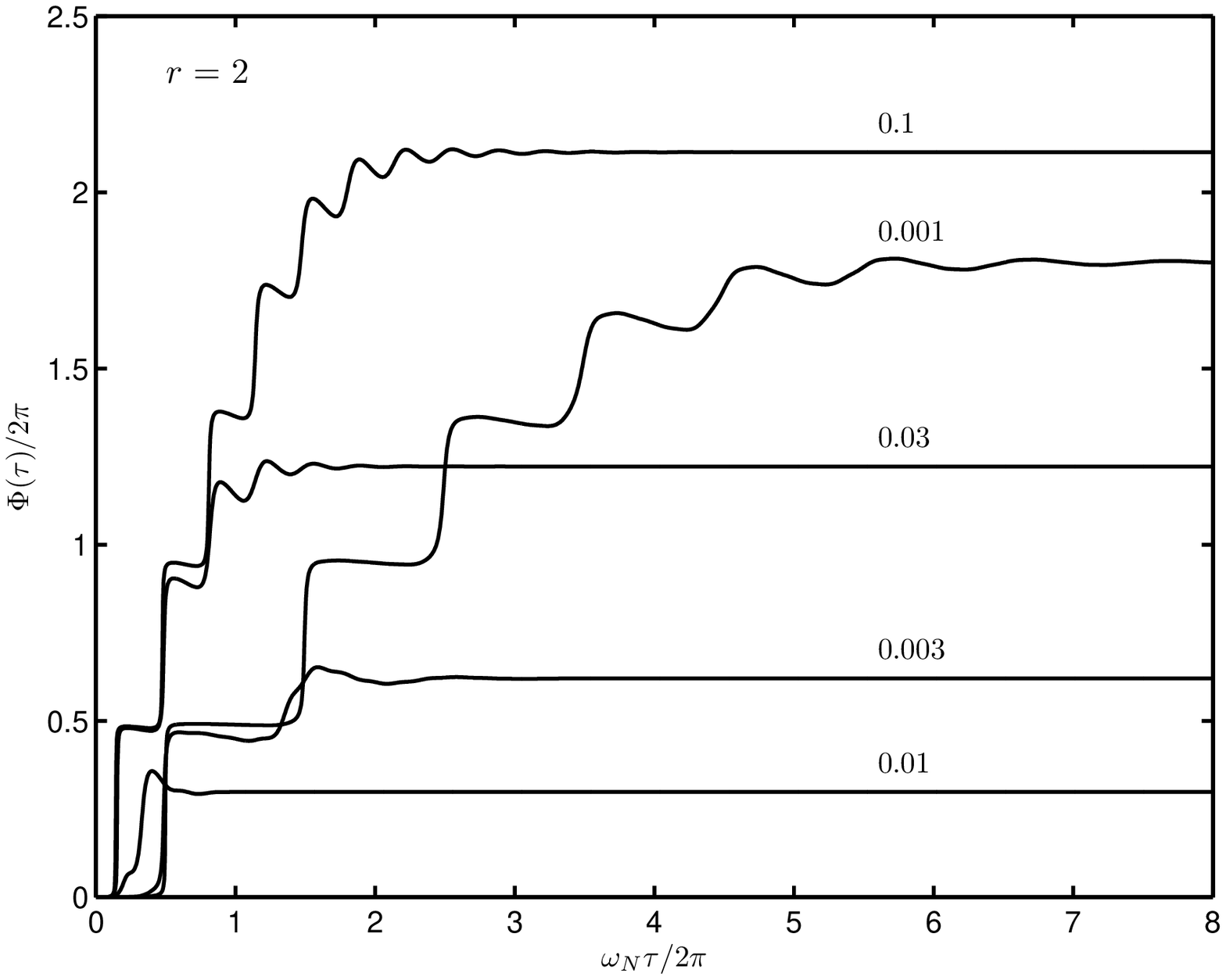}
\caption{The GP vs time for different coupling strengths for a
thermal environment (left) and a squeezed-thermal environment with
$r=2$ (right). The magnetic field is $\hbar \omega_N /k_BT=1$ and
the initial state is a pure spin up state along $x$. The values of
the ratio $g^2/\omega_N$ are shown on the graphs. As can be seen,
squeezing enhances the GP for a particular case with large coupling.
Note also small oscillations which are produced by the effect of
squeezed field on the spin state.} \label{fig:sq_g}
\end{figure}

\begin{figure}[H]
\includegraphics[scale=0.42]{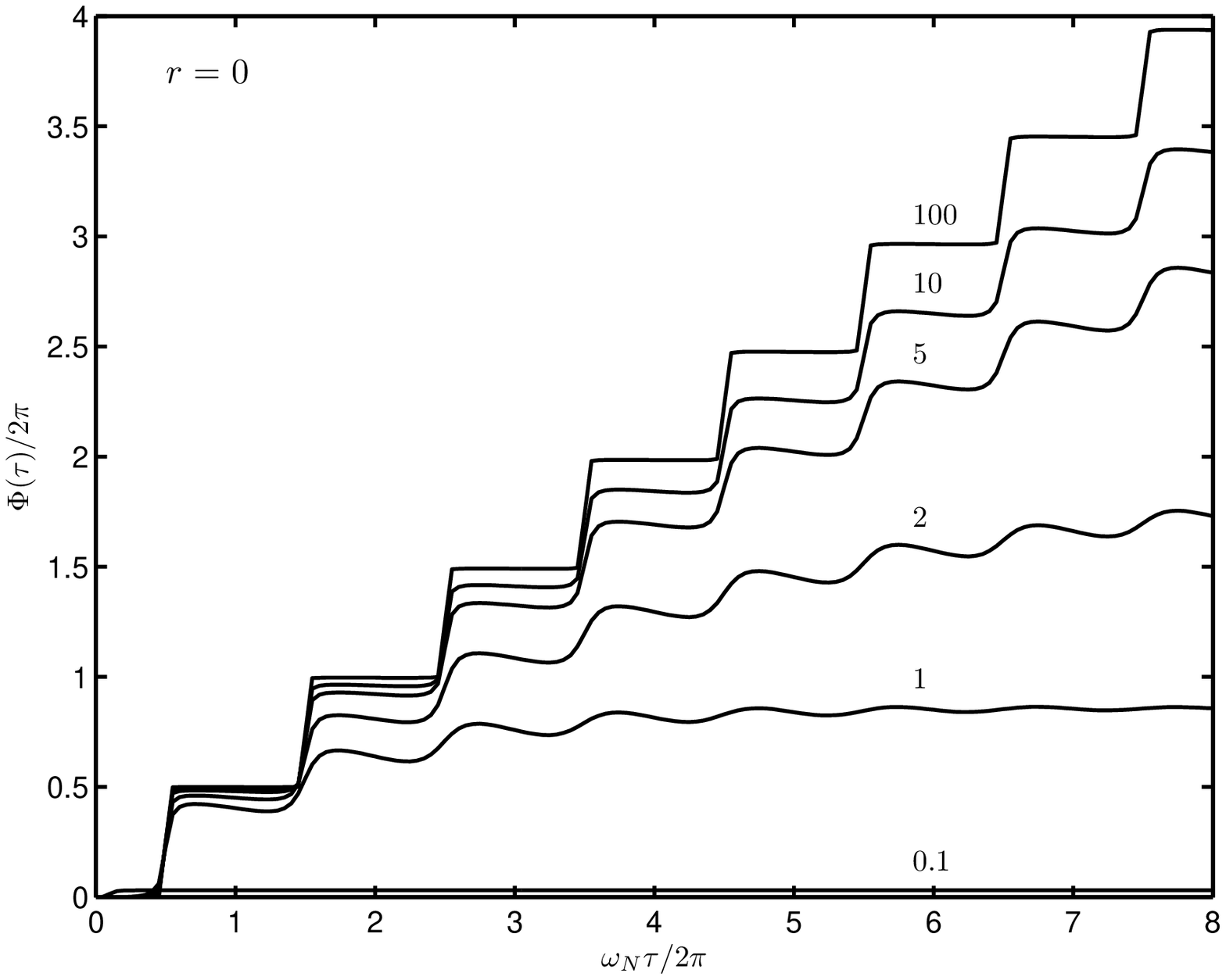}
\includegraphics[scale=0.42]{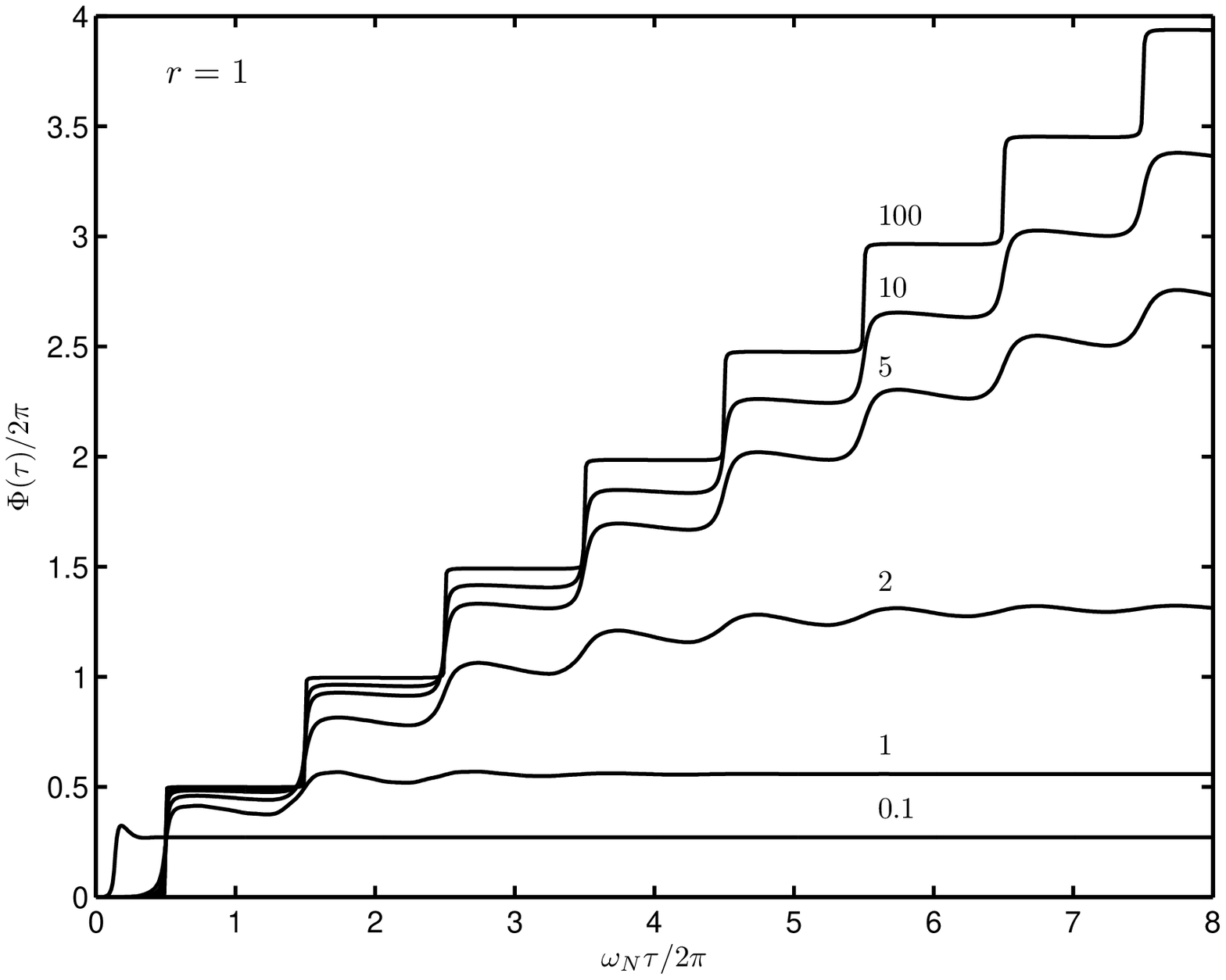}
\caption{The GP vs time for different magnetic fields for a thermal environment (left)
and a squeezed-thermal environment with $r=1$ (right).
The coupling coefficient is taken as $\hbar g^2/k_BT=0.01$ and the initial state is a pure spin up state along
$x$. The values of the ratio $\hbar\omega_N/k_BT$ are shown on the graph.
We have taken $\Omega=3\omega_N$ for all data points.}
\label{fig:sq_B}
\end{figure}

The effects of changing the temperature, the coupling strength and
the magnetic field on the time-dependence of the GP are shown in the
Figures \ref{fig:sq_temp}, \ref{fig:sq_g} and \ref{fig:sq_B},
respectively. As it can be seen, increasing the temperature or the
coupling strength, or decreasing the magnetic field have regular
effects when there is no squeezing (the figures on the left of
Fig.s~\ref{fig:sq_temp}, \ref{fig:sq_g} and \ref{fig:sq_B}). With
these changes, the oscillations of the GP are destroyed and the
limiting value of GP is reached at earlier times.

But when there is squeezing, that regular dependence is lost for the
case of increasing the temperature or the coupling strength
(Fig.s~\ref{fig:sq_temp} and \ref{fig:sq_g}). For sufficiently large
values of the temperature or the coupling constant, the limiting
value of the GP is higher in comparison to those with lower
temperatures or coupling strengths. Moreover the GP reaches the
limiting values at a shorter time for higher temperatures and at a
longer time for higher coupling strengths. For low temperatures and
coupling strengths, presence of squeezing in general decreases the
GP.

For high magnetic fields, the squeezing has almost no effect
(Fig.~\ref{fig:sq_B}). At lower fields, the squeezing tends to
decrease the GP, destroys the oscillations and make the limiting
values reached at earlier times. And for small enough field
strengths it is possible to get higher values for the GP with
increasing the squeezing.

\begin{figure}[h!]
\includegraphics[scale=0.42]{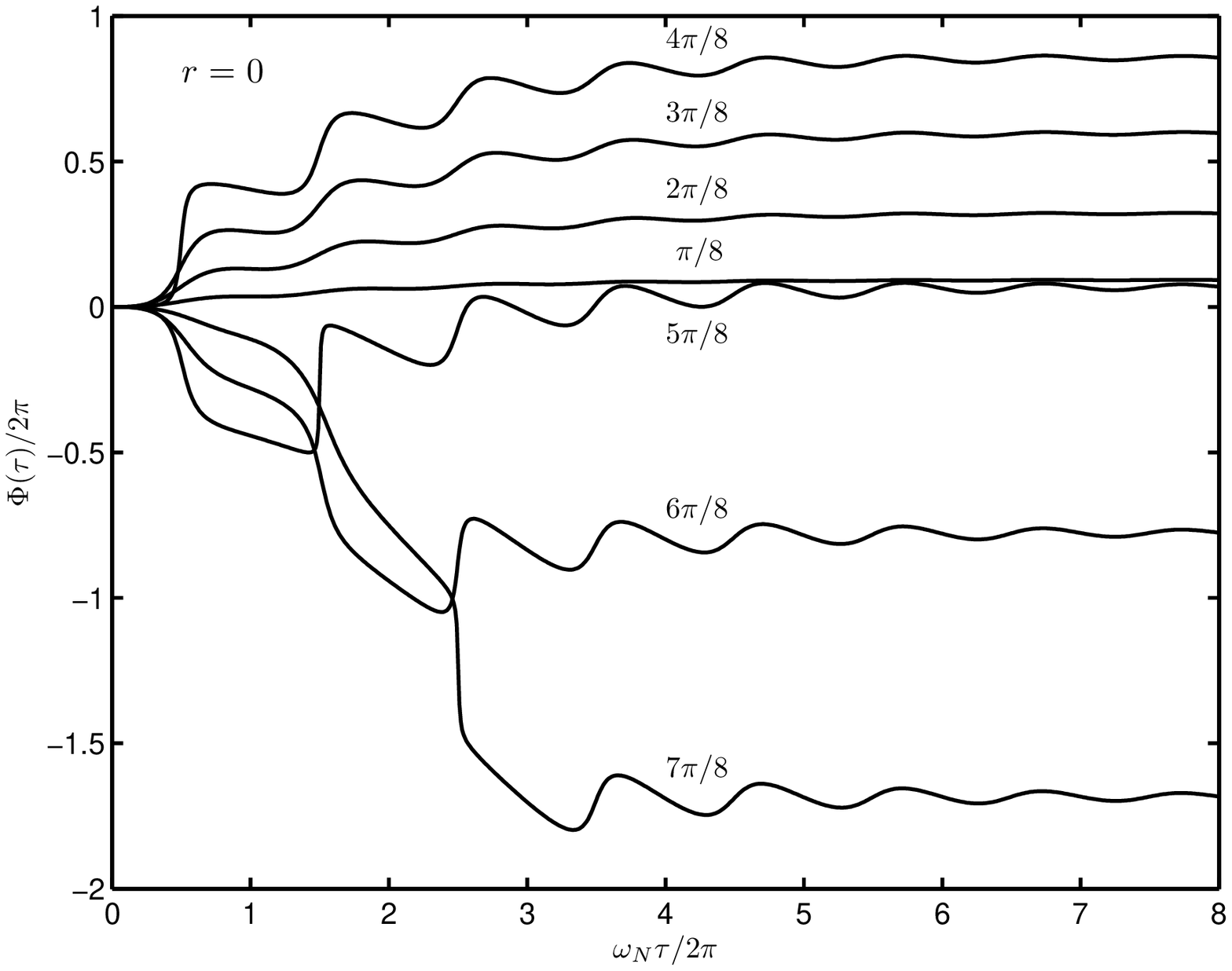}
\includegraphics[scale=0.42]{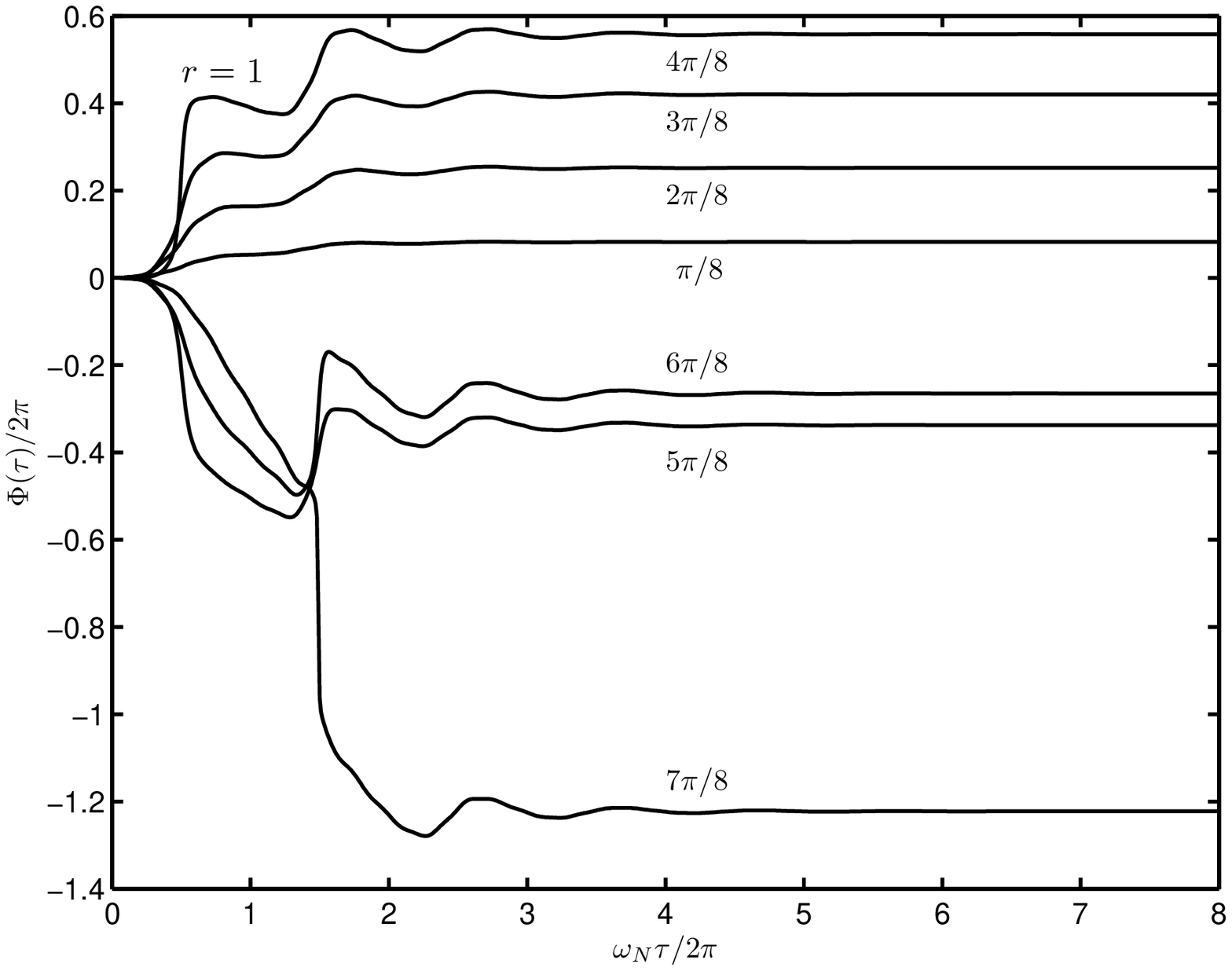}
\caption{The GP vs time for different initial states for a thermal environment (left)
and a squeezed-thermal environment with $r=1$ (right).
The magnetic field is $\hbar \omega_N /k_BT=1$ and the
coupling strength is $g^2/\omega_N=0.01$. The initial state is chosen as a pure state
and the value of the spherical angle $\theta$ of the initial
orientation is shown on the graph.} \label{fig:th_theta}
\end{figure}

Figure \ref{fig:th_theta} shows the dependence of the GP on the
initial state. The initial state is taken as a pure state along a
direction on the $xz$-plane having the spherical angle $\theta$.
When $\theta$ is increased above $\pi/2$, the GP changes
significantly. Especially, the GP tends to decrease at the initial
moments for $\theta>\pi/2$. If the initial state is mixed, then
the GP is suppressed in magnitude but its overall behavior does
not change.

\subsection{The Dependence of the Phase at Infinity on Physical
Parameters}

\begin{figure}[h!]
\includegraphics[scale=0.61]{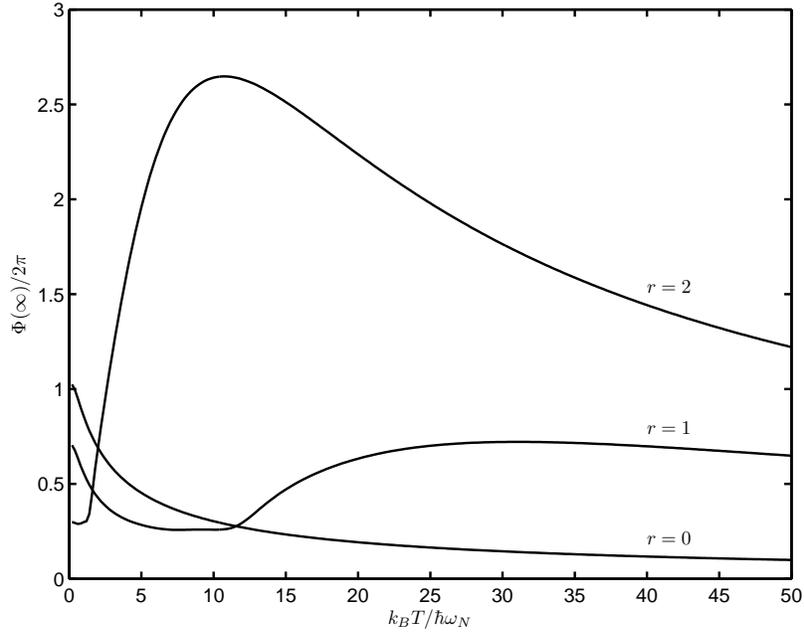}
\caption{The GP at infinity as a function of temperature for
different values of the squeezing parameter $r$. The coupling
strength satisfies $g^2/\omega_N=0.01$.} \label{fig:inf_temp}
\end{figure}

\begin{figure}[h!]
\includegraphics[scale=0.61]{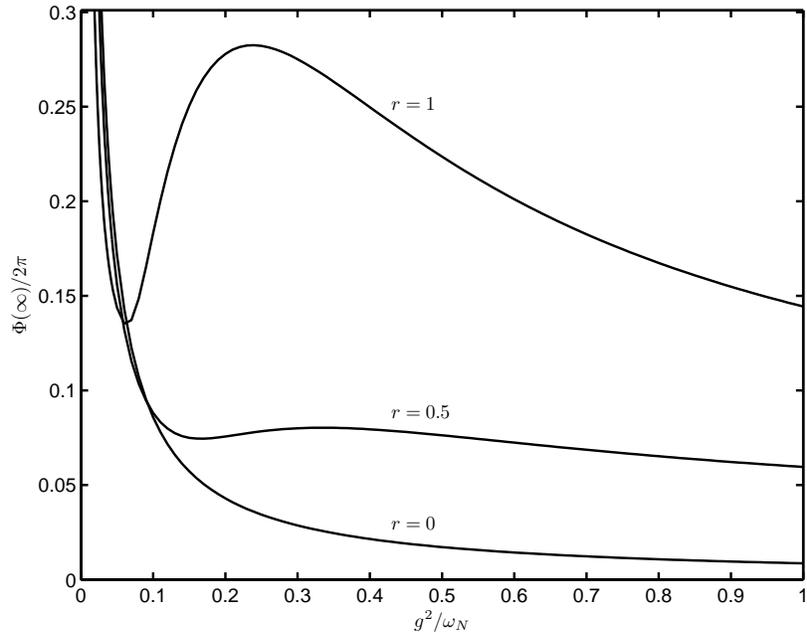}
\caption{The GP at infinity as a function of the coupling strength
for different values of the squeezing parameter $r$ for the case
$\hbar\omega_N/k_BT=1$. } \label{fig:inf_g}
\end{figure}

The behavior of the GP at infinity is also a quantity of interest.
Its magnitude gives an indication of the overall magnitudes of the
GP that can be obtained at any time. This value can also be used
for showing the effect of the squeezing on the GP. In
figure~\ref{fig:inf_temp}, the long time limit of the GP, i.e.,
$\Phi(\infty)$, is shown as a function of temperature. When there
is no squeezing, increasing the temperature always suppresses the
GP. With the presence of squeezing, three different temperature
regimes appear. For very low and very high temperatures, the GP is
still suppressed with increasing $T$. However, there is now an
intermediate range of temperatures over which the opposite trend
appears. Increasing the squeezing parameter $r$ moves this
intermediate range to lower temperatures. Fig.~\ref{fig:inf_g}
shows the dependence of the limiting value of the GP on the
coupling strength. The behavior is similar to the case of changing
the temperature parameter.

\begin{figure}[h!]
\includegraphics[scale=0.61]{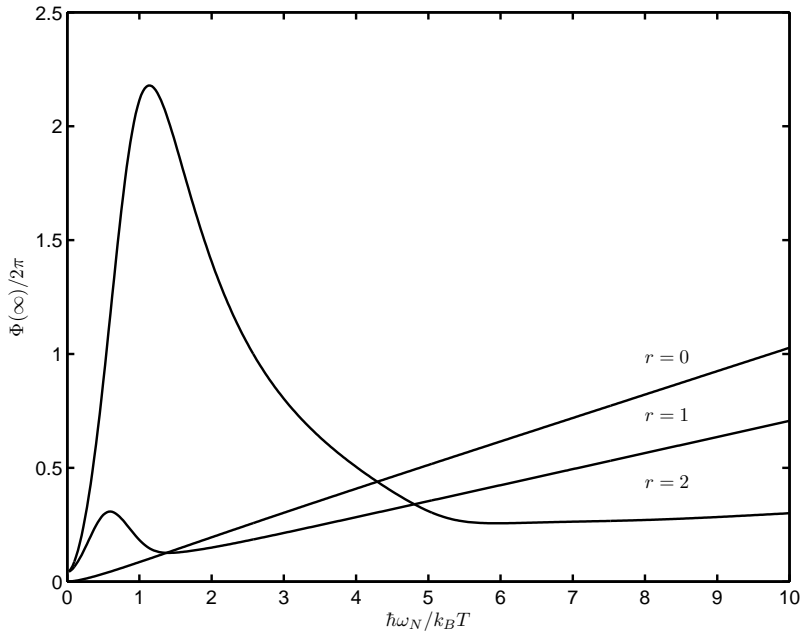}
\caption{The GP at infinity as a function of the magnetic field for
different values of the squeezing parameter $r$. The coupling
strength satisfies $\hbar g^2/k_BT=0.1$.} \label{fig:inf_B}
\end{figure}

Figure~\ref{fig:inf_B} shows the dependence of the limiting value of
the GP on the magnetic field. It can be seen that there are two
regions for this case. For high enough magnetic fields this
dependence is almost linear. Squeezing essentially decreases the
slope. But at lower magnetic fields, GP has a different behavior
under squeezing. In this regime, squeezing enhances the GP which is
now a nonlinear function of the magnetic field. The limiting GP also
reaches to a maximum value at an optimum value of the magnetic
field. Increasing the squeezing expands that region of nonlinearity
to higher magnetic fields.

\begin{figure}[H]
\begin{center}
\includegraphics[scale=0.61]{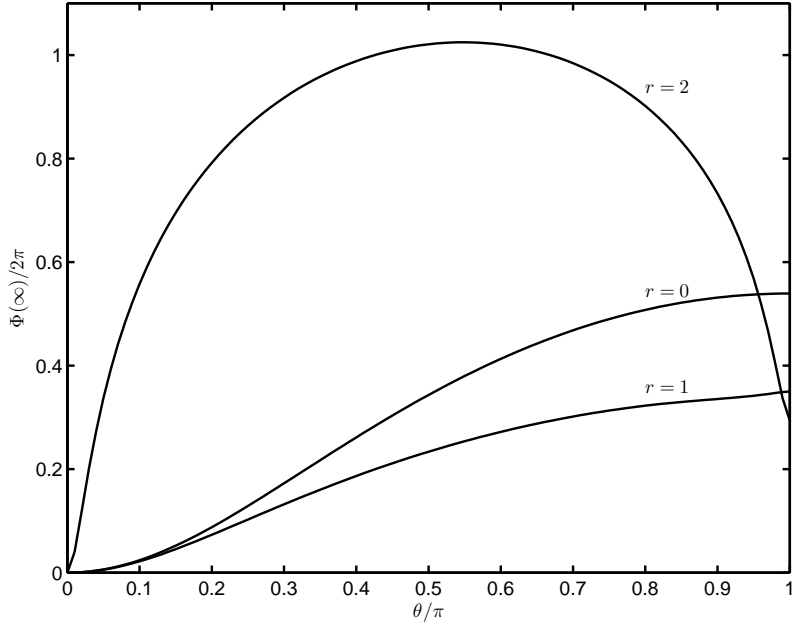}
\caption{The GP at infinity as a function of the initial state
angle $\theta$ for different values of the squeezing parameter $r$.
The parameters are chosen as $g^2/\omega_N=0.025$ and
$\hbar\omega_N/k_BT=1$.}
\end{center}
\label{fig:inf_theta}
\end{figure}

The dependence of $\Phi(\infty)$ on the initial state is shown in
Fig.~\ref{fig:inf_theta}. As can be seen, for low squeezing
$\Phi(\infty)$ reaches its largest values for $\theta$ near $\pi$,
i.e., a spin orientation which is almost spin down. Small
squeezing suppresses the GP but does not change the behavior. For
large squeezing however, $\Phi(\infty)$ can be enhanced
significantly and its largest values are reached when $\theta$ is
around $\pi/2$.

\section{Conclusion}
\label{sec:conc}

The availability of easy manipulation with the current technology
NMR techniques makes nuclear systems good candidates as carrier
systems of the entities that are necessary for quantum information
processing\cite{Ekert00,Barenco951,Cory97,Cory98,Jones98,Jones00}.
With this motivation, the effects of the coupling between a nucleus
and a dissipative environment on the GP that the state of the
nucleus gains during its time evolution are studied. Although it is
affected by the physical parameters, it is always possible to get a
finite GP. Its dependence on these parameters are almost as
expected; it decreases with increasing temperature or coupling
strength, because both increase the decoherence caused by the
environment, and it increases with increasing magnetic field, since
increasing $B$ has the same effect as decreasing both the
temperature and the coupling constant.

If the GP needs to be increased further, the environment could be
driven by an electromagnetic field in a squeezed state. However, in
this case, one should be very careful because the squeezing may have
a destructive or an enhancing effect on the GP depending on the
parameters. In order to enhance the GP, the temperature should be
held at some appropriate intermediate value; it should not be too
small or too large. The coupling strength should be taken at a
relatively high value so that the squeezed field can interact with
the spin more; but it should not be too large lest the spin relaxes
to its equilibrium state quickly. And larger magnetic fields do not
always enhance the GP; it is true only up to an upper bound for the
field strength. A geometric phase that can be described by the model
used in this article can promote the robustness in quantum
computations.


\end{document}